\documentclass[sn-mathphys-num]{sn-jnl}% Math and Physical Sciences Numbered Reference Style 
\usepackage{graphicx}%
\usepackage{multirow}%
\usepackage{amsmath,amssymb,amsfonts}%
\usepackage{amsthm}%
\usepackage{mathrsfs}%
\usepackage[title]{appendix}%
\usepackage{xcolor}%
\usepackage{textcomp}%
\usepackage{manyfoot}%
\usepackage{booktabs}%
\usepackage{algorithm}%
\usepackage{algorithmicx}%
\usepackage{algpseudocode}%
\usepackage{listings}%
\theoremstyle{thmstyleone}%
\theoremstyle{thmstyletwo}%

\theoremstyle{thmstylethree}%

\begin{document}

\title[Again About Singularity Crossing In Gravitation And Cosmology]{Again About Singularity Crossing In Gravitation And Cosmology}


\author*[1,2]{\fnm{Alexander} \sur{Kamenshchik}}\email{kamenshchik@bo.infn.it}



\affil*[1]{\orgdiv{Department of Physics and Astronomy ``A. Righi''}, \orgname{University of Bologna}, \orgaddress{\street{via Irnerio 46}, \city{Bologna}, \postcode{40126}, %\state{State}, 
\country{Italy}}}

\affil[2]{\orgdiv{Section of Bologna}, \orgname{INFN}, \orgaddress{\street{viale Berti Pichat 6/2 }, \city{Bologna}, \postcode{40127}, %\state{State}, 
\country{Italy}}}

%\affil[3]{\orgdiv{Department}, \orgname{Organization}, \orgaddress{\street{Street}, \city{City}, \postcode{610101}, \state{State}, \country{Country}}}


\abstract{We discuss the problem of singularity crossing in isotropic and anisotropic universes. We study at which conditions  singularities can disappear in quantum cosmology and how  quantum particles behave in the vicinity of singularities. Some  attempts to develop general approach to the connection between the field reparametrization and the elimination of singularities is presented as well.}




\maketitle

\section{Introduction}\label{sec1}

The problem of cosmological singularities has been attracting the attention of theoreticians
working in gravity and cosmology since the early 1950s \cite{Lif-Khal}. In the 1960s  general theorems
about the conditions for the appearance of singularities were proven \cite{Pen-Hawk,Pen} and the oscillatory
regime of approaching the singularity \cite{BKL}, called also ``Mixmaster universe'' \cite{Misner}, was discovered.
An amusing relation between infinite-dimensional hyperbolic Kac-Moody algebras \cite{Kac} and the chaotic approach to the singularity 
in various cosmological theories was discovered at the beginning of the new millennium \cite{DHN,Bel-Hen-book}. 

Basically, until the end of 1990s almost all discussions about singularities were devoted to the
Big Bang and Big Crunch singularities, which are characterized by a vanishing cosmological
radius.
However, kinematical investigations of Friedmann cosmologies have raised the possibility
of sudden future singularity occurrence \cite{sudden}, characterized by a diverging  acceleration $\ddot{a}$, whereas both the
scale factor $a$ and  $\dot{a}$  are finite. Then, the Hubble parameter $H = \frac{\dot{a}}{a}$  and the energy density $\rho$
are also finite, while the first derivative of the Hubble parameter and the pressure $p$ diverge.
Until recent years, however, the sudden future singularities attracted rather a limited interest
of researchers.

The discovery of  cosmic acceleration \cite{cosmic,cosmic1} stimulated the development of ``exotic'' cosmological models of dark energy \cite{dark};  in some of these models  sudden (soft)  future singularities arise quite naturally.
Thus, during last years a plenty of work, devoted to the study of future singularities was done (see, for example, the review \cite{my-review} and references therein). 
A distinguishing feature of these singularities is the fact that rather often they can be crossed \cite{Fern,Fern1}. Indeed, due to the finiteness of the Christoffel symbols, the geodesic equations are not singular and, hence, can be continued through these singularities without difficulties.   
Here a new and interesting phenomenon arises:
in some models interplay between the geometry and  matter forces  matter to change some of its basic properties, such as the equation of state for fluids and even the form of  the Lagrangian for fields \cite{paradox1,my-review}.

Naturally, the regularity of geodesics  does not mean that the objects existing in the universe can pass through the singularities without substantial changes or even destruction. To study this question one should consider non only the geodesic equations, including the Christoffel symbols, but also the geodesic deviation equation, where the components of the Riemann-Christoffel curvature tensor are present. Some aspects of the behaviour of the quantum particles in the vicinity of singularities will be reviewed in the section 4 of the present paper.

Generally, the crossing of the Big Bang -- Big Crunch singularity is more complicated and looks more counter-intuitive with respect to the crossing of the soft future singularities. However, during last years some approaches to this problem were elaborated \cite{Bars,Bars1,Wetterich,Wetterich1,Prester,Prester1}. Behind of these approaches there are basically two general ideas. Firstly, to cross the singularity means to give a prescription matching non-singular, finite quantities before and after such a crossing. Secondly, such a description can be achieved by using a convenient choice of the field parametrization. In our paper \cite{we-cross} we have developed a method of the description of the singularity crossing where we have used the technique of the transition between the Einstein and Jordan frames \cite{frame,frame1}. Our results concerning the soft singularities crossing, the transformations of the matter properties and the description of the Big Bang--Big Crunch crossing in the Friedmann-Lema\^itre universes were presented at the first Lema\^itre conference in 2017 and then were published in the  paper \cite{Found}. Here I would like to describe the results obtained after the first Lema\^itre conference 
(and also some results in quantum cosmology, which were not presented in 2017) which were discussed in my talk at the second Lema\^itre conference in 2024.
The structure of  the paper is as follows: the second section is devoted to anisotropic cosmological models in the presence of a scalar field; in the third section we discuss the relation between quantum cosmology and the problem of cosmological singularities; in the fourth section we analyse the behaviour of quantum particles in the vicinity of cosmological singularities; the fifth section is devoted to the description of attempts to elaborate a general covariant approach to the problem of the singularity crossing; 
in the sixth section we consider a Bianchi-I universe filled with a spatially homogenous magnetic field; the last section contains concluding remarks.

\section{Anisotropic cosmological models, scalar fields, gravity and singularity crossing}\label{sec2}

In our  paper \cite{we-cross} we proposed a version of the
description of the crossing of singularities in universes filled
with scalar fields. This version was based on the transitions
between the Jordan and the Einstein frames. Conformal
symmetry also was used in our approach because we
implemented a particular choice of the coupling between
the scalar field and the scalar curvature--the conformal
coupling.We used a conformal coupling because in this case
the relations between the parametrizations of the scalar field
in the Jordan and in the Einstein frames have a simple explicit
form. For simplicity we only considered an isotropic
cosmological singularity, present in a flat Friedmann universe.
We also essentially used the relations between exact
solutions of the cosmological (Friedmann and Klein-Gordon) equations in two different frames, which were
studied in detail in our papers \cite{frame,frame1}. The main idea of
Ref. \cite{we-cross} was the following: when in the Einstein frame the
universe arrives at the Big Bang--Big Crunch singularity, from
the point of view of the evolution of its counterpart in the
Jordan frame, its geometry is regular, but the effective
Planck mass has  a zero value. The solution to the equations
of motion in the Jordan frame is smooth at this point, and by
using the relations between the solutions of the cosmological
equations in two frames, one can describe the crossing of the
cosmological singularity in a uniquely determined way. The
contraction is replaced by the expansion (or vice versa), and
the universe enters into the antigravity regime. Analogously,
when the geometry is singular in the Jordan frame, it is
regular in the Einstein frame, and using this regularity, we
can describe in a well-determined way the crossing of the
singularity in the Jordan frame.

Let us note that the method of the description of the singularity crossing described above, is not connected directly with the much debated question of the equivalence between 
the Jordan frame and the Einstein frame (see e.g. \cite{EF,EF1,EF2,EF3,EF4,EF5,EF6}).  Briefly speaking, we think that mathematically these frames are equivalent, but physically the cosmological evolutions connected by the transformations between the frames can be rather different and exactly this fact was used in the procedure described above.  

The method, implemented in \cite{we-cross} does not work in anisotropic  universes. To see the reason it is enough to consider the simplest Bianchi-I universe, 
as it was done in paper \cite{we-cross1}. Let us suppose that a scalar field $\phi$, nonminimally coupled to gravity is present in this universe.
The metric of the Bianchi-I universe in the Jordan frame has the following form:
\begin{equation*}
d\tilde{s}^2 = \tilde{N}(\tau)^2d\tau^2-\tilde{a}^2(\tau)(e^{2\beta_1(\tau)}dx_1^2+e^{2\beta_2(\tau)}dx_2^2+e^{2\beta_3(\tau)}dx_3^2),
\end{equation*}
where the anisotropic factors $\beta_i$ satisfy the relation
\begin{equation*}
\beta_1+\beta_2+\beta_3=0.
\end{equation*}
The transition to the Einstein frame gives the metric 
\begin{equation*}
ds^2 = N(\tau)^2d\tau^2-a^2(\tau)(e^{2\beta_1(\tau)}dx_1^2+e^{2\beta_2(\tau)}dx_2^2+e^{2\beta_3(\tau)}dx_3^2).
\end{equation*}
Then, analysing the equations of motion in the Jordan frame, one has 
\begin{equation*}
\dot{\beta}_i=\frac{\beta_{i0}}{\tilde{a}^3}, \theta_0=\beta_{10}^2+\beta_{20}^2+\beta_{30}^2.
\end{equation*}
\begin{equation*}
\dot{\phi} = \frac{\phi_0}{\tilde{a}^3},\ \phi = \frac{\phi_0}{\left(\frac{3\theta_0}{2}+\frac{3\phi_0^2}{4U_1}\right)^{\frac12}}\ln \tilde{t},
\end{equation*}
where 
\begin{equation*}
\theta_0=\beta_{10}^2+\beta_{20}^2+\beta_{30}^2
\end{equation*}
and $U_1$ is an arbitrary (positive) constant arising at the transition  from the Jordan frame to the Einstein frame.  
In the vicinity of the singularity in the Einstein frame 
\begin{equation*}
\tilde{a} \sim \tilde{t}^{\frac{1}{3}}.
\end{equation*}
In the Jordan frame 
\begin{equation*}
a \sim  \tilde{t}^{\frac{1}{3}}(\tilde{t}^{\gamma}+\tilde{t}^{-\gamma}) \rightarrow 0,
\end{equation*}
because 
\begin{equation*}
\gamma= \frac{\phi_0}{3\sqrt{\phi_0^2+2\theta_0U_1}} < \frac13.
\end{equation*}
Thus, one also encounters the Big Bang singularity  in the Jordan frame. Note, that in the case of isotropic universe, $\theta_0 = 0, \gamma = \frac13$ and the 
singularity in the Jordan frame disappears. Thus, to treat the singularity in the anisotropic universes one should look for another method. 
We shall try it first coming back to the Friedmann-Lema\^itre model.

The Friedmann-Lema\^{i}tre model with a massless scalar field can be described by the Lagrangian
\begin{equation*}
L = \frac12\dot{x}^2-\frac12\dot{y}^2,
\end{equation*}
where
\begin{equation*}
x = \frac{4\sqrt{U_1}}{\sqrt{3}}\tilde{a}^{\frac32} \cosh \frac{\sqrt{3}}{4\sqrt{U_1}}\phi,\ y= \frac{4\sqrt{U_1}}{\sqrt{3}}\tilde{a}^{\frac32} \sinh \frac{\sqrt{3}}{4\sqrt{U_1}}\phi,
\end{equation*}
and the Friedmann equation is 
\begin{equation*}
\dot{x}^2-\dot{y}^2=0.
\end{equation*}
Inversely,
\begin{equation*}
\tilde{a}^3 = \frac{3(x^2-y^2)}{16U_1},
\end{equation*}
\begin{equation*}
\phi=\frac{4\sqrt{U_1}}{\sqrt{3}}{\rm arctanh}\frac{x}{y}.
\end{equation*}
Initially our new variables satisfy an inequality 
\begin{equation*}
x>|y|.
\end{equation*}
The solution is
\begin{equation*}
x=x_1\tilde{t}+x_0,\ y=y_1\tilde{t}+y_0,\ x_1^2=y_1^2.
\end{equation*}
Choosing the constants as 
\begin{equation*}
x_0=y_0=A>0,\ x_1=-y_1 = B>0,
\end{equation*}
we have
\begin{equation*}
\tilde{a}^3=\frac{3AB\tilde{t}}{4U_1}.
\end{equation*}
We can make a continuation in the plane $(x,y)$ to $x < |y|$ or, in other words, to  $\tilde{t} < 0$.
Such a continuation implies an antigravity regime and the transition to the phantom scalar field, just as in the more complicated schemes, discussed before (see \cite{we-cross} and the references therein). 

How can we generalize these considerations to the case when the anisotropy term is present?
We can consider the following Lagrangian:
\begin{equation*}
L = \frac12\dot{r}^2-\frac{1}{2}r^2(\dot{\varphi}^2+\dot{\varphi}_1^2+\dot{\varphi_2}^2),
\end{equation*}
\begin{equation*}
\varphi_1=\sqrt{\frac38}\alpha_1,\ \varphi_2\sqrt{\frac38}\alpha_2,
\end{equation*}
\begin{equation*}
\beta_1=\frac{1}{\sqrt{6}}\alpha_1+\frac{1}{\sqrt{2}}\alpha_2,\ \beta_2=\frac{1}{\sqrt{6}}\alpha_1-\frac{1}{\sqrt{2}}\alpha_2,\ \beta_3=-\frac{2}{\sqrt{6}}\alpha_1.
\end{equation*}
We can again consider the plane $(x,y)$ as 
\begin{eqnarray*}
&&x = r\cosh \Phi,\\
&&y=r\sinh \Phi,
\end{eqnarray*} 
where a new hyperbolic angle $\Phi$ is defined by
\begin{equation*}
\Phi=\int d\tilde{t}\sqrt{\dot{\varphi}_1^2+\dot{\varphi}_2^2+\dot{\varphi}^2}.
\end{equation*}
We have reduced a four-dimensional problem to the old two-dimensional one, on using the fact that the variables $\alpha_1,\alpha_2$ and $\phi$ enter into the equation of motion for the scale factor $\tilde{a}$ only through the squares of their time derivatives.

The behaviour of the scale factor before and after the crossing of the singularity can be matched by using the transition to the new coordinates $x$ and $y$, which mix geometrical and scalar field variables in a particular way.
To describe the behaviour of the anisotropic factors it is enough to fix the constants $\beta_{i0}$.    Detailed analysis of the evolution of the Bianchi-I universe filled with massless scalar field including the singularity crossings was presented in the subsequent paper \cite{we-cross2}. Here the cases of the minimal coupling, conformal coupling and of the induced gravity \cite{induced,induced1} were considered and compared. 

The paper \cite{we-cross3} was devoted to the study of another family of the anisotropic cosmological models: to the Kantowski-Sachas models \cite{K-S} filled with a scalar field.  It is well-known that the geometry of the spacetime of the Kantowski-Sachs universe is connected with the Schwarzschild black hole geometry (see e.g. \cite{Schwarz-we} and the references therein). Indeed, under the horizon the time and radial cooordinates exchange their roles and the interior part of black hole represents a Kantowski-Sachs universe. However, there is also another kind of relation between the static solutions of the Einstein equations and time dependent cosmological solutions.   
A duality between  spherically symmetric
static solutions in the presence of a massless scalar field and the Kantowski-Sachs cosmological models
which instead possess  hyperbolic symmetry was found in paper \cite{we-dual} and further studied in papers \cite{we-dual1,we-cross3,we-dual2}. 
The main ingredient of this duality is the exchange of roles between the radial coordinate and the temporal coordinate combined with the exchange between the spherical two-dimensional  geometry and the hyperbolical two-dimensional geometry.
We have found exact solutions for static spherically and hyperbolically symmetric geometries in the presence of a massless scalar field conformally coupled to gravity and their respective  Kantowski-Sachs cosmologies \cite{we-cross3}. These solutions are convenient for an analysis of the singularity crossing problem. 
Here we present an exact formula for the respective cosmological solutions, possessing two-dimensional hyperbolic symmetry. The metric has the following form:
\begin{eqnarray*}
&&ds^2=\frac{a_0^2\left(A_0\left(\tanh\frac{t}{2}\right)^{2\sqrt{\frac{1-\gamma^2}{3}}}+1\right)^2\sinh^2t}{4A_0\left(\tanh\frac{t}{2}\right)^{2\gamma+2\sqrt{\frac{1-\gamma^2}{3}}}}
\times
(dt^2-d\chi^2-\sinh^2\chi d\phi^2)\\
&&-\frac{b_0^2\left(A_0\left(\tanh\frac{t}{2}\right)^{2\sqrt{\frac{1-\gamma^2}{3}}}+1\right)^2\left(\tanh\frac{t}{2}\right)^{2\gamma-2\sqrt{\frac{1-\gamma^2}{3}}}}{4A_0}dr^2,
\label{KS}
\end{eqnarray*}
where the coefficient $\gamma$ characterises an intensity of the scalar field present in the universe. 
For a particular value $\gamma = 1/2$ this metric acquires the following form:
\begin{eqnarray*}
&&ds^2=\frac{a_0^2\left(A_0\tan\frac{t}{2}+1\right)^2\cos^4 \frac{t}{2}}{A_0}
(dt^2-d\theta^2-\sin^2\theta d\phi^2)
-\frac{b_0^2\left(A_0\tan\frac{t}{2}+1\right)^2}{4A_0}dr^2.
\label{KS4}
\end{eqnarray*}
This metric is regular at $t=0$ and has  singularities at $t=\pm \pi$ and at $t = t_0 = -2{\rm arctan}\frac{1}{A_0}$.

At $t \rightarrow  \pm \pi$ the scale factor $a \rightarrow 0$ while $b \rightarrow \infty$. At $t \rightarrow t_0$ both  scale factors vanish.

At $t < 0$, we find ourselves in the region with antigravity because $U_c < 0$. 
The expression  contains only integer powers of the trigonometrical functions and one can describe  the crossing of the singularities in a unique way.
Thus, we can imagine an infinite periodic evolution of the universe. 

In the vicinity of the moment $t \rightarrow \pi$, the asymptotic expressions for the metric coefficients are 

\begin{equation*}
ds^2 = dT^2 -c_1^2Td\theta^2 -c_2^2T\sin^2\theta d\phi^2 - c_3^2\frac{1}{T}dr^2.
\label{Kasner}
\end{equation*}
This form has a structure similar to that of the Kasner solution for a Bianchi-I universe,
where the Kasner indices have the values

\begin{equation*}
p_1=\frac12,\ p_2 = \frac12,\ p_3=-\frac12.
\label{Kasner1}
\end{equation*}
These indices do not satisfy the standard Kasner relations 

\begin{equation*}
p_1+p_2+p_3=p_1^2+p_2^2+p_3^2=1,
\label{Kasner2}
\end{equation*}
they satisfy the generalized relation

\begin{equation*}
\sum _{i=1}^3p_i^2=2\sum_{i=1}^3p_i-\left(\sum_{i=1}^3p_i\right)^2.
\label{Kasner3}
\end{equation*}
In the vicinity of the singularity at $t=t_0$: 
\begin{equation*}
ds^2 = dT^2 -c_1^2Td\theta^2 -c_2^2T\sin^2\theta d\phi^2 - c_3T dr^2.
\label{Kasner4}
\end{equation*}
 This behavior is isotropic. 
 
 Let us note that the cases $\gamma = \pm 1$ correspond to the empty universes and their description is well known.
 On the other hand, it is not yet clear how can one treat the cases $\gamma \neq \pm 1, \gamma \neq \pm \frac12$ and how 
 the singularity crossing can be described. Perhaps, this question is worth of further investigations.

\section{Quantum cosmology and singularities}\label{sec3}

Speaking about quantum cosmology and singularities, researchers usually  mean one of two different things:
One of the approaches  is based on the idea that the consideration of quantum corrections to the classical General Relativity implies some modification  of the Friedmann equation:
\begin{equation*}
\frac{\dot{a}^2}{a^2}+\frac{k}{a^2}=\rho_{\rm matter} +\rho_{\rm quantum\ corrections}.
\end{equation*}
Such a modification can, in turn, imply the appearance of solutions free of cosmological singularity, which sometimes are called ``bounces''. 

Another, more fundamental approach, is based on the study of the Wheeler-DeWitt equation \cite{DeWitt}, which defines the quantum state of universe, or, the wave function of the universe. The Wheeler-DeWitt equation arises as a result of the application of the Dirac quantisation procedure for the constrained systems to the universe as a whole and has the following form: 
\begin{equation*}
\hat{\cal H}\Psi = 0,
\end{equation*}
where $\Psi$ is the wave function of the universe, and $\hat{\cal H}$ is the operator, obtained by means of the quantisation of the so called super-Hamiltonian constraint. 
Already in his original paper \cite{DeWitt}, DeWitt has suggested that the vanishing of the wave function of the universe at the singularity means that the quantum theory eliminates the classical singularity. This approach, however, requires some precisations. The point is that the Wheeler-DeWitt equation is not only very complicated. There are two other interconnected problems behind  its simple appearance in an abstract form. The first problem is connected with the  so called question of time in quantum cosmology (see e.g. \cite{Kiefer} and the references therein). Indeed, the time parameter is absent in the Wheeler-DeWitt equation. Second problem consists in the fact that the solution of the Wheeler-DeWitt equation does not have an immediate probabilistic interpretation. These problems can be treated in a combined way by choosing of some gauge condition explicitly dependent on time.
Then, the time is identified with some combination of the phase variables of the theory. For the rest of the variables one obtains an effective  Schr\"odinger equation. On doing this one should define in a consistent way the scalar product on the reduced Hilbert space, taking into account the presence of the Faddeev-Popov determinant \cite{Barv,Barv-Kam}. 
When all this done, one can associate singularities with such values of the variables in the reduced Hilbert space, which correspond to the appearance of singularities in classical theory.

The analysis of the Wheeler-DeWitt equation in the presence of  soft future singularities was undertaken in paper \cite{Barbara}. There a particular kind of the future singularity -- 
``Big Brake'' was studied. The Big Brake singularity is such a singularity when in finite cosmic time the universe arrives to the state with a finite value of the scale factor, zero value of the velocity of expansion and an infinite value of the deceleration parameter. The Big Brake was discovered in paper \cite{we-tach}, where a particular cosmological model with a tachyon field \cite{Sen,Padman, Feinstein,tach-Chap} was introduced. Note, that what was called ``tachyon'' field is nothing but some kind of the Born-Infeld type of scalar field \cite{Born-Infeld}. 

Let us describe our model \cite{we-tach} briefly.
The tachyon Lagrange density in a flat Friedmann universe is 
\begin{eqnarray*}
L = - V(T)\sqrt{1-\dot{T}^2},
\end{eqnarray*}
its  energy density is
\begin{eqnarray*}
\rho =  \frac{V(T)}{\sqrt{1-\dot{T}^2}}
\end{eqnarray*}
and the pressure is 
\begin{eqnarray*}
p = - V(T)\sqrt{1-\dot{T}^2}.
\end{eqnarray*}
The equation of motion for the tachyon field is
\begin{equation*}
\frac{\ddot{T}}{1-\dot{T}^{2}}+3H\dot{T}+\frac{V_{,T}}{V}=0.
\end{equation*}
In our model \cite{we-tach} the potential has a rather particular form
\begin{eqnarray*}
&&V(T)=\frac{\Lambda }{\sin ^{2}\left[ \frac{3}{2}{\sqrt{\Lambda \,(1+k)}\ T}%
\right] }
\times\sqrt{1-(1+k)\cos ^{2}\left[ \frac{3}{2}{\sqrt{\Lambda \,(1+k)}\,T}%
\right] }\ ,  \label{VTfixed}
\end{eqnarray*}
where $k$ and $\Lambda > 0$ are the parameters of the model.
The case $k > 0$ is more interesting. Two unusual phenomena here take place: the transformation of the tachyon field into another kind of the 
Born-Infeld type field - ``pseudotachyon'' and appearance of the Big Brake singularity. The model \cite{we-tach} was further studied in \cite{we-tach01,we-tach1,paradox,paradox1} and some other interesting phenomena, like the transformation of matter properties \cite{paradox1} were found. However, the Big Brake arises in a much more simple model of the universe filled with a perfect fluid which was called \cite{we-tach} ``anti-Chaplygin gas'' and which has an equation of state
\begin{equation*}
 p = \frac{A}{\rho},
 \end{equation*}
 where $A$ is a positive constant. 
 
 Another, relatively simple model, where the Big Brake arises is the model with a minimally coupled scalar field with the potential, which is negative and inversely proportional 
 to the value scalar field
 \begin{equation*}
V(\phi) = -\frac{B}{\phi}.
\end{equation*}
This model was studied in \cite{Barbara}, where it was shown that exact solutions in the vicinity of the singularity can be written down and the relative wavefunction disappears.

In paper \cite{Serena} the results concerning the scalar field model were confirmed and the models with tachyon field were also considered. It was shown that while the 
wavefunctions of the universe tend to zero at the singularity, the probability density does not vanish because, taking into account the Faddeev-Popov determinant contribution, one sees that it includes a compensating factor. In other words, the quantum probability of  finding the universe crossing Big Brake singularity is nonvanishing and this singularity is not ``suppressed'' quantistically. 

At the same time rather simple considerations show that the Big Bang -- Big Crunch singularity is suppressed in quantum cosmology, at least, in some of its models.
Let us consider, for example, the universe filled with scalar field and require its normalizability \cite{we-norm}. In the reduced Hilbert space the wave function depends on this scalar field and requirement of the normalizability is  
\begin{equation*}
\int d\phi \bar{\Psi}(\phi)\Psi(\phi) < \infty.
\end{equation*}
To guarantee this inequality, one has to require that 
when $|\phi| \rightarrow \infty$, the probability density $\bar{\Psi}\Psi$ should tend to zero rapidly.
If $|\phi| \rightarrow \infty$ corresponds to Big Bang -- Big Crunch singularity, then this singularity is suppressed due to the requirement of the normalizability of the wave function of the universe. In the subsequent paper \cite{Nick} a more detailed analysis of classical and quantum cosmology of some Born-Infeld type models was undertaken and the results of the preceding papers were confirmed.

\section{Particles, fields and singularities}\label{sec4}

The question what happens with particles (in quantum field theoretical sense) when the universes passes through the cosmological singularity was studied in \cite{Olesya}. 
All the considerations were carried out for a  scalar field in a flat Friedmann universe satisfying the Klein-Gordon equation:
\begin{equation*}
\Box\phi +V'(\phi) = 0.
\label{K-G}
\end{equation*} 
One can consider a spatially homogeneous solution of this equation $\phi_0$, depending only on time $t$ as a classical background.
A small deviation from this background solution can be represented as a sum of  Fourier harmonics satisfying linearised equations
\begin{equation*}
\ddot{\phi}(\vec{k},t)+3\frac{\dot{a}}{a}\dot{\phi}(\vec{k},t)+\frac{\vec{k}^2}{a^2}\phi(\vec{k},(t)) + V''(\phi_0(t))\phi(\vec{k},(t)) = 0.
\label{K-G1}
\end{equation*}
The corresponding quantised field is 
\begin{equation*}
\hat{\phi}(\vec{x},t) = \int d^3\vec{k}(\hat{a}(\vec{k})u(k,t)e^{i\vec{k}\cdot\vec{x}}+\hat{a}^+(\vec{k})u^*(k,t)e^{-i\vec{k}\cdot\vec{x}}),
\label{quant}
\end{equation*}  
where the creation and the annihilation operators satisfy the standard commutation relations:
\begin{equation*}
[\hat{a}(\vec{k}), \hat{a}^+(\vec{k}')]=\delta(\vec{k}-\vec{k}').
\label{quant1}
\end{equation*}
The basis functions should be normalised 
\begin{equation*}
u(k,t)\dot{u}^*(k,t)-u^*(k,t)\dot{u}(k,t)=\frac{i}{(2\pi)^3a^3(t)},
\label{canon2}
\end{equation*}
so that the canonical commutation relations between the field $\phi$ and its canonically conjugate momentum 
 $\hat{{\cal P}}$ 
\begin{equation*}
[\hat{\phi}(\vec{x},t), \hat{{\cal P}}(\vec{y},t')]=i\delta(\vec{x}-\vec{y})
\label{canon}
\end{equation*}
were satisfied.

The linearised Klein-Gordon equation has two independent solutions.
 To define a  particle it is necessary to have two independent non-singular solutions.
 It is a non-trivial requirement in the situations when a singularity or other kind of irregularity of the spacetime geometry occurs.
 It will be  convenient also to construct explicitly the vacuum state for quantum particles as a Gaussian function of the corresponding variable. Let is introduce an operator
\begin{equation*}
\hat{f}(\vec{k},t) = (2\pi)^3(\hat{a}(\vec{k})u(k,t)+\hat{a}^+(-\vec{k})u^*(k,t)).
\label{vac-exp}
\end{equation*}
Its canonically conjugate momentum is 
\begin{equation*}
\hat{p}(\vec{k},t) = a^3(t)(2\pi)^3(\hat{a}(\vec{k})\dot{u}(k,t)+\hat{a}^+(-\vec{k})\dot{u}^*(k,t)).
\label{vac-exp1}
\end{equation*}
We can express the annihilation operator as 
\begin{equation*}
\hat{a}(\vec{k}) =i\hat{p}(\vec{k},t)u^*(k,t)-ia^3(t)\hat{f}(\vec{k},t)\dot{u}^+(k,t).
\label{vac-exp2}
\end{equation*}
Representing the operators $\hat{f}$ and $\hat{p}$ as 
\begin{equation*}
\hat{f} \rightarrow f,\ \ \hat{p} \rightarrow -i\frac{d}{df},
\label{vac-exp3}
\end{equation*}
one can write down the equation for the corresponding vacuum state in the following form:
\begin{equation*}
\left(u^*\frac{d}{df}-ia^3\dot{u}^*f\right)\Psi_0(f) = 0.
\label{vac-exp4}
\end{equation*}
Then
\begin{equation*}
\Psi_0(f) = \frac{1}{\sqrt{|u(k,t)|}}\exp\left(\frac{ia^3(t)\dot{u}^*(k,t)f^2}{2u^*(k,t)}\right).
\label{vac-exp5}
\end{equation*}

Now, we are in a position to study the behaviour of quantum fields and particles in the vicinity of  cosmological singularities of different types substituting the corresponding evolution laws for scale factors into the equations presented above.  

In the case of the Big Bang - Big Crunch singularity, one of the basis functions in the vicinity of the singularity becomes singular and it is impossible to construct a Fock space.

In the case of the Big Rip singularity, when in finite interval of time the universe achieves an infinite volume and infinite time derivative of the scale factor, the Fock space can be constructed for a spectator scalar field, but it does not exist for the phantom scalar field driving the expansion.

In the case of the model with tachyon field \cite{we-tach}, discussed in the preceding section, we have considered three situations.
The first situation regards the non-singular transformation of the tachyon into the pseudo-tachyon field with the Lagrangian
\begin{equation*}
L = W(T)\sqrt{\dot{T}^2-1}.  
\end{equation*}
In this case both basis functions are regular  and hence 
the operators of creation and annihilation are well defined. 
However, at the  moment of the transformation the dispersion of the Gaussian wave function of the vacuum becomes infinite and then becomes finite again.
Then, we analyse what happens in the vicinity of the Big Brake singularity and see that there  it is impossible to define a Fock vacuum.
Then we consider a more complicated model \cite{paradox1}, where the  dust is added to the Born-Infeld type field. Here the character of the soft singularity is slightly changed and then the presence of the Fock vacuum is restored in the vicinity of the soft singularity.

We have already mentioned above the cosmological models where a universe, driven by phantom scalar field (i.e. the field with a negative kinetic term), arrives to a Big Rip singularity. For such an evolution a particular behaviour of the equation of state parameter $w = \frac{p}{\rho}$ is typical. Namely, $w < -1$. During some period researchers were also interested in cosmological models where for some part of the evolution $w > -1$ and for some - $w <-1$. The value $w=-1$ was called ``phantom divide line'' and the corresponding phenomenon was called ``phantom divide line crossing''. It is easy to observe this phenomenon in the models with two scalar fields - one normal and one phantom. It is more interesting to study models where a unique scalar field changes its nature. Such models were studied in \cite{phan-cross,phan-cross1}. 
A characteristic feature of these models is that the potential of the scalar field has a cusp.
Remarkably, a passage through
the point where the Hubble parameter achieves a maximum value implies the change of the sign of the kinetic
term. Though a cosmological singularity is absent in
these cases, this phenomenon is a close relative of those,
considered in the preceding section, because here we also
find some transformation of matter properties induced by
a change of the geometry. In this aspect the phenomenon of
the phantom divide line crossing in the model \cite{phan-cross,phan-cross1}
is analogous to the transformation between the tachyon
and pseudotachyon field in the Born-Infeld model with
the trigonometric potential discussed earlier.
Let us add that the potential with a cusp in \cite{phan-cross,phan-cross1}
has a form
\begin{equation*}
V(\phi) = \frac{V_0}{(1+V_1\phi^{\frac32})^2}.
\end{equation*}

Analysing the dynamics of the model in the vicinity of the phantom divide line, we have seen that in this case both the solutions of the corresponding 
linearised equation for the perturbations are regular, but both of them tend to zero. What does it mean?
It means that the  dispersion of the Gaussian
function, describing the vacuum state of the scalar field  tends to zero and the function becomes the Dirac
delta function. After the crossing of the phantom divide line the dispersion becomes regular again. One can interpret this as follows:
for a moment the vacuum and the Fock space disappear
and then their reappear once again, while the particles of
the phantom field become the particles of the standard scalar field with a positive kinetic term. In spite of an apparently unusual character 
of this behaviour of the wave packet it is quite logical for the transition from one scalar field to another (phantom) having different signs of the kinetic terms. 

\section{Covariant approach to the singularities}\label{sec5}

The crossing of the Big Bang - Big Crunch singularities looks rather counterintuitive.
However, as it was discussed before, it can be sometimes described by using the reparametrization of fields, including the metric.
One can say that to do this, it is necessary to resort to one of two ideas, or a combination
thereof. One of these ideas is to employ a reparameterisation of the field variables which makes
the singular geometrical invariant non-singular.
Another idea is to find such a parameterisation of the fields, including, naturally, the metric,
that gives enough information to describe consistently the crossing of the singularity even
if some of the curvature invariants diverge.
The application of these ideas looks in a way as a  craftsman work. 

   Our goal in papers \cite{Ibere,Ibere1,Ibere2} was to develop a general formalism to distinguish ``dangerous'' and ``non-dangerous'' singularities,
studying the field variable space of the model under consideration and trying to answer the question:
When the spacetime singularities can be removed by a reparametrisation of the field variables?
We have tried to elaborate the following hypothesis: when the geometry of the space of the field variables is non-singular, the elimination of curvature singularities is possible. 

Our treatment of the  field space $\mathcal S$ was inspired by the concept of the Vilkovisky-DeWitt effective action \cite{Vilk,DeWitt1}. 
The goal was  to treat on the same (geometrical) footing both changes of coordinates in the spacetime $\mathcal M$ and field redefinitions in the functional approach to quantum field theory.
This approach requires introducing a local metric $\boldmath G$ in field space $\mathcal S$ and computing
the associated geometric scalars by defining a covariant derivative which is compatible with $\boldmath G$.
The metric $\boldmath G$ is actually determined by the kinetic part of the action and its dimension depends
on the field content of the latter.

After some cumbersome calculations in the functional space, we have shown that the Kretschmann scalar 

\begin{equation*}
\mathcal K = \mathcal R_{ABCD}\,\mathcal R^{ABCD}
\end{equation*}
is finite in every theory of pure gravity
\begin{equation*}
\mathcal K =\frac{n}{8}\left(\frac{n^3}{4} + \frac{3 n^2}{4} - 1\right),
\end{equation*}
 where $n$ is the spacetime dimension. 
 This result can be interpreted as a statement that all the singularities in empty universe can be crossed. 

We have introduced another method working in the functional space, based on the notion of the  quantum effective action and study of its topological characteristics \cite{Ibere1}. Let us discuss it briefly.
We introduce the functional
\begin{equation*}
\psi[\varphi]
=
e^{i\,\Gamma[\varphi]}
\end{equation*}
and we shall call $\psi[\varphi]$ the functional order parameter because $\psi$ plays
the analogous role of an order parameter in the theory of phase transitions in ordered media
or cosmology (see e.g. \cite{Mermin}).

The field space $\mathcal M$ can be thought of as the ordered medium itself,
whereas functional singularities correspond to topological defects. 
The functional order parameter $\psi$ defines the map
\begin{equation*}
\psi : \mathcal M \to \mathbb S^1,
\end{equation*}
from the field space to the unit circle, the latter playing the role of the
order parameter space.
The singularities can be characterized by the fundamental group (first homotopy group).
Since $\pi_1(\mathbb S^1) = \mathbb Z$, the homotopy classes are labeled by the
winding number $\mathcal W$. 
A functional singularity exists whenever $\mathcal W \neq 0$.

We have considered as an example  a flat Friedmann universe filled with a massless scalar field, minimally coupled to gravity. 
In this model there is the singularity of the Big Bang - Big Crunch type. 
This singularity can be eliminated by a field reparametrisation (see the papers \cite{we-cross,we-cross1}) discussed before.  
Direct (while tricky) calculation shows that in this case the winding number is indeed equal to zero \cite{Ibere1}, as we had expected.

\section{Bianchi-I cosmologies, magnetic fields and singularities}\label{sec6}

The examples and approaches discussed in the preceding sections were based on the consideration of an empty universe or a universe filled with different 
(standard or non-standard) scalar fields. Thus, it makes sense to dwell on another example: the universe in the presence of magnetic fields. It is well-known that  this
problem represents not only mathematical interest. 
The existence of large-scale magnetic fields in our universe is an important and
enigmatic phenomenon (see, e.g,~Refs.~\cite{Thorne,Barrow-magn,numbers,numbers0}).
Their origin is not known and is being widely discussed.
Thus, even if it is unrealistic to describe the present-day universe with the Bianchi-I metric, models, where this metric is sustained
by a magnetic field, could shed some light on processes that occurred in the very early universe.
While the literature on the universes filled by magnetic field is rather vast, we thought that it was useful to try to reconsider the case of the Bianchi-I universe filled with the spatially homogeneous magnetic field oriented along one of its axes using modern language and connecting it with the problem of the singularity crossing \cite{Polina}. 

If we represent the Bianchi-I metric in the form
\begin{equation*}
ds^2 = dt^2 -(a^2(t)dx^2+b^2(t)dy^2+c^2(t)dz^2)
\end{equation*}
and suppose that the magnetic field is oriented along the axis $z$, then the energy density of this magnetic field will have form 
\begin{equation*}
\rho = \frac{B_0^2}{a^2b^2},
\end{equation*}
while its pressure will be anisotropic: its $z$-component will be negative, $x$ and $y$ components will be positive and their absolute values coincide with that of the energy density. Analysing the corresponding Einstein equations and the Maxwell equations, we find that in the beginning of cosmological evolution such a universe finds itself in a Kasner regime \cite{Kasner}, and, moreover, the Kasner indices satisfy the standard Kasner relations and can be conveniently parameterised  by the Lifshitz-Khalatnikov parameter $u$ as follows 
\cite{Lif-Khal}:
\begin{eqnarray*}
&&p_1 = -\frac{u}{1+u+u^2},\\
&&p_2 = \frac{u+1}{1+u+u^2},\\
&&p_3 = \frac{u(u+1)}{1+u+u^2}.
\end{eqnarray*}
Then, studying the behaviour of the universe in the long distant future when its volume $abc$ tends to infinity, we find it again in the Kasner regime, but the values of the Kasner indices are changed and, respectively, is changed the value of the parameter $u$. Remarkably, the law of the changes of the Kasner indices and of the $u$-parameter 
coincides with that which occurs in the Bianchi-II universe \cite{Taub} and which is responsible for the oscillatory approach to the cosmological singularity \cite{BKL}.   
Namely, if the initial value of the Lifshitz-Khalatnikov parameter $u > 2$, then we have the transition:
\begin{equation*}
u \rightarrow u-1,
\end{equation*}
which in paper \cite{BKL} was called `` change of Kasner epoch''. If, instead, $u < 2$, then the operation $u \rightarrow u-1$ should be followed by another transformation 
\begin{equation*}
u \rightarrow \frac{1}{u},
\end{equation*}
which was called in \cite{BKL} ``change of Kasner era''. The fact that in the presence of the magnetic field the universe in the vicinity of the singularity lives in a Kasner regime means that  for the description of the singularity crossing we can use the methods described in \cite{we-cross1,we-cross2}.

\section{Conclusions}\label{sec7}

The problem of singularities in gravitation and cosmology attracts an  attention of researchers in the field not only because of its mathematical and theoretical importance.
It is also connected with such aspects of the life of the Universe as its origin and its future fate, which could be relevant also for persons whose work and life are not directly connected to physics. The papers and books touching these ``existential'' aspects of the singularities are rather numerous and here we can cite only few of them
\cite{Star-future,Kam-future,Cycles,Mercati,Yurov,Odin1,Mercati1,Odin} (naturally, our choice is rather subjective). 

In this paper we have presented a brief review of works, which were unified by a general idea, that it is reasonable to try to co-exist with singularities and to try to integrate them into a ``normal'' theory. However, the idea that it would be better to have a theory and/or  particular models totally free of singularities conserves its attractiveness for many researchers. The flow of works devoted to attempts to construct regular black holes or nonsingular universes is rather intensive.   As examples of recent papers where the non-singular cosmologies were studied, we can cite \cite{modern,modern1,modern2,we-reg-cosm,Kam-Pet1}. Very active is also the search for regular black hole solutions. The study of non-singular black holes started a long time ago~\cite{Bardeen}; ( for recent reviews, see~\cite{Spallucci,Sebastiani}). Recently, there was a new burst of activity in this field, starting with the paper \cite{Simpson-Visser}.  One can write down a singularity-free metric ansatz from the Schwarzschild black hole by a simple substitution of the radial coordinate $r$ as $r \rightarrow \sqrt{r^2+b^2}$, as was proposed by Simpson and Visser~\cite{Simpson-Visser}. That results in the following spacetime:
\begin{equation*}
ds^2 = \left(1-\frac{2m}{\sqrt{r^2+b^2}}\right)dt^2-\left(1-\frac{2m}{\sqrt{r^2+b^2}}\right)^{-1}dr^2-\bigl(r^2+b^2\bigr)d\Omega_2^2, %(d\theta^2+\sin^2\theta d\varphi^2),
\label{S-V} 
\end{equation*}
 where $b$ is a parameter and the singularity at $r=0$ is replaced by a regular minimum of the area function at $r=0$, a sphere of radius $b$. If $b > 2m$, this spacetime represents a wormhole with a throat at $r=0$; if $b <2m$, one has a black hole with two horizons at $r=\pm\sqrt{4m^2-b^2}$; and the $b=2m$ case corresponds to an extremal black hole with a single horizon at $r=0$. At the hypersurface $r=0$ in the black hole case, the coordinates change their temporal and spatial assignments, which corresponds to a bounce in one of the two scalar factors of the Kantowski-Sachs universe, the so-called black bounce. Some other papers, using the algorithm similar to that used  in the paper \cite{Simpson-Visser} and studying different regular black hole geometries were published during the last couple of years \cite{Vaidya,charged,Kerr,Bron,Bronnikov,Kam-Pet}. 
 
 Generally, in the majority of works devoted to the construction of regular black holes, one  uses the method, which many years ago was called by Synge ``G-method'' \cite{Synge}; (see also a recent e-print \cite{Ellis}). Using this method, one chooses a metric, substitutes it into the left-hand side of the Einstein equations, and then sees what happens on the right-hand side. The G-method is opposed to the ``T-method'', when one chooses the form of the matter in the right-hand side of the Einstein equations and then tries to find the metric that satisfies this system of equations, being substituted into their left-hand side. The advantage of the G-method consists in the fact that it always works (in contrast with the T-method). Unfortunately, the right-hand side of the Einstein equations, arising as the result of the application of the G-method, does not always have some reasonable physical sense and can be identified with some known fields or other types of matter. The remarkable example of the regular black hole sustained by the minimally coupled phantom scalar field with explicitly known potential was found in~\cite{Bron-Fabris}. Some properties of this solution were studied in further detail~\cite{we-reg-cosm}, and it was also used \cite{Kam-Pet} as an attempt to construct a regular rotating black hole by using Newman-Jannis algoritm \cite{N-J}.  However, one can say that 
 almost all regular black holes or universes can hardly be sustained by a reasonable kind of matter in contrast to such classical singular solutions like Schwarzschild black hole or Friedmann-Lema\^itre universe. Thus, one can say that there is some kind of complementarity between the regularity of the geometry of the solutions of the Einstein equations and the simplicity (or naturalness) of its matter content. From our point of view one is not obliged to try to avoid singularities by any means and the rational treatment of them could represent a valid direction of research. 
 
Concluding this paper we would like to make reference to a remarkable paper by Misner \cite{Misner1}, published as early as in 1969. There the idea that the singularities in the General Relativity are not its drawback but its distinguishing feature, which should be accepted, and the adequate language for their treatment should be developed, is expressed in a very clear and convincing way. In particular, he wrote, ``We should stretch our minds, find some more acceptable set of words to describe the mathematical situation, now identified as ``singular", and then proceed to incorporate this singularity into our physical thinking until observational difficulties force revision on us.
The concept of a true initial singularity (as distinct from an indescribable early era at extravagant but finite high densities and temperatures) can be a positive and useful element in cosmological theory.''

\bmhead{Acknowledgements}
I am grateful to the organisers of the Second International conference to celebrate the Legacy of G. Lema\^itre. Black Holes, Gravitational Waves and Space-Time Singularities (June 17-21, 2024, Castel Gandolfo, Specola Vaticana), for an invitation to give a talk and a partial financial support.

\end{document}